\documentclass[12pt]{article}

\usepackage{amsmath,array,amssymb,cite} 

\def\elabel#1{\label{#1}}

\DeclareSymbolFont{AMSb}{U}{msb}{m}{n}
\DeclareSymbolFontAlphabet{\Bbb}{AMSb}

\def\rsen{\setcounter{equation}{0}}

\def\N{{\cal N}}

\def\tr{{\rm tr}}

\def\dalpha{{\dot\alpha}}

\def\hf{{\textstyle{1\over2}}}

\def\sigmabar{\bar\sigma}

\def\N{{\cal N}}

\def\lambdabar{\bar\lambda}

\def\sst{\scriptscriptstyle}

\def\D{{\cal D}}
\def\Dbarslash{\,\,{\raise.15ex\hbox{/}\mkern-12mu {\bar\D}}}
\def\delslash{\,\,{\raise.15ex\hbox{/}\mkern-9mu \partial}}
\def\Dslash{\,\,{\raise.15ex\hbox{/}\mkern-12mu \D}}

\def\sigmabar{\bar\sigma}
\def\psibar{\bar\psi}

\def\dalpha{{\dot\alpha}}

\def\N{{\cal N}}

\def\hf{{\textstyle{1\over2}}}

\def\hf{{\textstyle{1\over2}}}

\def\N{{\cal N}}

\def\dalpha{{\dot\alpha}}

\def\sigmabar{\bar\sigma}

\def\VEV#1{\left\langle #1\right\rangle}
\def\Vev#1{\big\langle{#1}\big\rangle}

\def\sigmabar{\bar\sigma}

\def\N{{\cal N}}

\def\hf{{\textstyle{1\over2}}}

\def\dalpha{{\dot\alpha}}

\def\lambdabar{\bar\lambda}
\def\psibar{\bar\psi}

\def\chitilde{\tilde{\chi}}

\def\uA{\,\lower 1.2ex\hbox{$\sim$}\mkern-13.5mu A}
\def\uphi{\,\lower 1.2ex\hbox{$\sim$}\mkern-13.5mu \phi}
\def\uX{\,\lower 1.2ex\hbox{$\sim$}\mkern-13.5mu X}
\def\uD{\,\lower 1.2ex\hbox{$\sim$}\mkern-13.5mu {\rm D}}

\def\uF{\,\lower 1.2ex\hbox{$\sim$}\mkern-13.5mu F}
\def\uW{\,\lower 1.2ex\hbox{$\sim$}\mkern-13.5mu W}
\def\uWbar{\,\lower 1.2ex\hbox{$\sim$}\mkern-13.5mu {\overline W}}

\def\uV{\,\lower 1.2ex\hbox{$\sim$}\mkern-13.5mu V}
\def\uv{\,\lower 1.0ex\hbox{$\scriptstyle\sim$}\mkern-11.0mu v}
\def\uPsi{\,\lower 1.2ex\hbox{$\sim$}\mkern-13.5mu \Psi}
\def\uPhi{\,\lower 1.2ex\hbox{$\sim$}\mkern-13.5mu \Phi}
\def\uchi{\,\lower 1.5ex\hbox{$\sim$}\mkern-13.5mu \chi}
\def\Psibar{\bar\Psi}
\def\uPsibar{\,\lower 1.2ex\hbox{$\sim$}\mkern-13.5mu \Psibar}
\def\upsi{\,\lower 1.5ex\hbox{$\sim$}\mkern-13.5mu \psi}
\def\psibar{\bar\psi}
\def\upsibar{\,\lower 1.5ex\hbox{$\sim$}\mkern-13.5mu \psibar}
\def\upsibarzero{\,\lower 1.5ex\hbox{$\sim$}\mkern-13.5mu \psibar^\zero}
\def\ulambda{\,\lower 1.2ex\hbox{$\sim$}\mkern-13.5mu \lambda}
\def\ulambdabar{\,\lower 1.2ex\hbox{$\sim$}\mkern-13.5mu \lambdabar}
\def\ulambdabarzero{\,\lower 1.2ex\hbox{$\sim$}\mkern-13.5mu \lambdabar^\zero}
\def\ulambdabarnew{\,\lower 1.2ex\hbox{$\sim$}\mkern-13.5mu \lambdabar^\new}
\def\D{{\cal D}}

\def\N{{\cal N}}
\def\Dslash{\,\,{\raise.15ex\hbox{/}\mkern-12mu \D}}
\def\Dbarslash{\,\,{\raise.15ex\hbox{/}\mkern-12mu {\bar\D}}}
\def\delslash{\,\,{\raise.15ex\hbox{/}\mkern-9mu \partial}}
\def\delbarslash{\,\,{\raise.15ex\hbox{/}\mkern-9mu {\bar\partial}}}

\begin{document}

\addtolength{\baselineskip}{4pt}
\thispagestyle{empty}

\begin{flushright}
{\tt hep-th/9905015}\\
May 1999
\end{flushright}


\begin{center}
{\scshape\Large Gluino Condensate
and Magnetic Monopoles\\ 
\vspace{0.15cm}
in Supersymmetric Gluodynamics\\}

\vspace{1cm}

{\scshape N. Michael Davies$^{1}$, Timothy J.~Hollowood$^{2,3}$,
Valentin V.~Khoze$^1$\\
and Michael P.~Mattis$^2$}

\vspace{0.2cm}
$^1${\sl Department of Physics, University of Durham,\\
Durham, DH1 3LE, UK}\\
 {\tt n.m.davies@durham.ac.uk}, {\tt valya.khoze@durham.ac.uk}\\

\vspace{0.3cm}
$^2${\sl Theoretical Division T-8, Los Alamos National Laboratory,\\
Los Alamos, NM 87545, USA}\\
 {\tt pyth@schwinger.lanl.gov}, {\tt mattis@lanl.gov}\\

\vspace{0.2cm}
$^3${\sl Department of Physics, University of Wales Swansea,\\
Swansea, SA2 8PP, UK}\\

\vspace{1cm}

{\Large ABSTRACT}
\end{center}

\vspace{0.1cm}

\noindent We examine supersymmetric $SU(N)$
gauge theories on ${\Bbb R}^3 \times S^1$ with a circle of circumference $\beta$.
These theories interpolate between four-dimensional ${\cal N}=1$
pure gauge theory for $\beta=\infty$ and three-dimensional
${\cal N}=2$ gauge theory for $\beta=0$. The dominant field configurations of the 
${\Bbb R}^3 \times S^1$ $SU(N)$ theories in the semi-classical regime
arise from $N$ varieties of monopole. Periodic instanton configurations correspond
to mixed configurations of $N$ single monopoles of the $N$ different types.
We semi-classically evaluate the 
non-perturbatively generated superpotential of the ${\Bbb R}^3 \times S^1$ theory
and hence determine its vacuum structure. We then calculate monopole 
contributions to the gluino condensate in these theories and take the 
decompactification limit $\beta=\infty$. In this way we obtain a value for the
gluino condensate in the four-dimensional ${\cal N}=1$ supersymmetric
$SU(N)$ Yang-Mills theory,
which agrees with the previously known `weak coupling' expression but
not with the `strong coupling' expression 
derived in the early literature solely from instanton considerations.
Moreover, we discover that the superpotential gives a mass to the dual (magnetic) photon,
which implies confinement of the original electric photon and disappearance
of all the massless modes.
 
\newpage

\section{Introduction}

The main goal of this paper is to provide a new 
calculation of the value of the 
gluino condensate
in four-dimensional $\N=1$ supersymmetric pure $SU(N)$ gauge theory.
Our approach incorporates recent results and ideas of 
Refs.~\cite{HKLM,LY,KL,LL,KvBzer,KvBone,KvBtwo}.
Previous to this, two conceptually different approaches for calculating 
$\Vev{\tr \lambda^2\over16\pi^2}$ have been followed in the literature:

1. In the first methodology \cite{NSVZone,ARV,AKMRV}, the so-called
strong-coupling instanton (SCI) approach, the gluino condensate 
$\Vev{\tr \lambda^2\over16\pi^2}$
is determined directly in the strongly coupled theory
using an explicit one-instanton calculation of 
$\Vev{{\tr\lambda^2(x_1)\over16\pi^2}\cdots{\tr\lambda^2(x_N)\over16\pi^2}}$. 
Cluster decomposition arguments are then invoked in order to
extract $\Vev{\tr \lambda^2\over16\pi^2}$.

2. In the second methodology \cite{NSVZtwo}, 
the so-called weak-coupling instanton (WCI) approach,
the calculation is
performed with additional matter fields whose presence ensures that the theory
is weakly coupled and a semi-classical `constrained instanton' calculation
is justified \cite{ADS}.
Holomorphicity of supersymmetric F-terms is then used to decouple
the matter fields and to flow to the original pure $\N=1$ gauge theory.

As reviewed in \cite{HKLM}, 
these two methods give two different 
values for the gluino condensate \cite{NSVZtwo,FS,AKMRV,FP}:
\begin{subequations}\begin{align}
&\VEV{\tr \lambda^2\over16\pi^2}_{\rm SCI}\ =\ 
{2 \over [(N-1)! \ (3N-1)]^{1/N}} \ \Lambda^3
\ ,\elabel{stwka} \\
&\VEV{\tr \lambda^2\over16\pi^2}_{\rm WCI}\ =\   \Lambda^3
\ .\elabel{stwkb} 
\end{align}\end{subequations}
These results are quoted in the Pauli-Villars scheme
with $\Lambda$ being the corresponding dimensional transmutation scale
of the theory. The reason for the discrepancy between the SCI versus WCI
calculations, as well as the question as to which is correct, 
has been a long-standing controversy
\cite{NSVZtwo,AKMRV,KS,SVrev}.
The new ingredient in this old puzzle 
is the fact that over the last few years {\it multi\/}-instanton
technology has been developed \cite{MO,measone,meastwo,KMS,DHKMV} 
to the extent that 
calculations 
can be performed in supersymmetric (and in principle non-supersymmetric) 
gauge theories,
both in the weak-coupling 
\cite{MO} and in the strong-coupling regimes \cite{DHKMV}, providing us
with successful quantitative tests of, respectively, the Seiberg-Witten
solution of $\N=2$ theories \cite{SW} and the Maldacena duality
\cite{Maldacena} in the $\N=4$ theory.
In \cite{HKLM} we re-examined the gluino condensate controversy using 
these recently developed
methods. In particular, we evaluated the large $N$ contribution of
$k$ instantons to gluino correlation functions
and demonstrated conclusively that an essential technical step in the SCI 
calculation
of the gluino condensate, namely the use of cluster decomposition, 
is actually invalid.

The central idea pursued in the present paper
is that there are additional configurations 
which contribute to gluino condensate in the strongly-coupled regime,
implying that the instanton-induced SCI expression \eqref{stwka} is
incomplete. The existence of other contributions necessarily affects
the application of cluster decomposition in the following sense.
Since each single instanton has $2N$
adjoint fermion zero-modes the $k$-instanton configuration 
contributes to the correlation function
\begin{equation}
\VEV{{\tr\lambda^2(x_1)\over16\pi^2}\cdots{\tr\lambda^2(x_{kN})\over16\pi^2}}\
, 
\elabel{cork}\end{equation}
rather than {\it directly\/} to
$\Vev{\tr \lambda^2\over16\pi^2}$ itself. In the SCI approach
gluino condensate is obtained by applying cluster decomposition to 
\eqref{cork} with the additional assumption that the instanton calculation
averages over the $N$ physically equivalent vacua of the $\N=1$ theory.
In the following, we will show that 
the correlator \eqref{cork} is not dominated solely by 
instantons and hence the clustering argument must be applied to the complete 
expression and not just to the instanton contribution. This is of course
in complete agreement with our earlier observation \cite{HKLM}\ that 
clustering fails when only multi-instantons are included in the SCI calculation.
Furthermore, when the theory is partially compactified on 
${\Bbb R}^3\times S^1$,
we will identify the configurations contributing {\it directly\/} to
$\Vev{\tr \lambda^2\over16\pi^2}$ with monopoles. 
By considering the contribution of
the monopole configurations, we will be able to argue that the complete
strong coupling expression for gluino condensate  
is different from the SCI expression \eqref{stwka} but agrees perfectly with
the WCI result \eqref{stwkb}.

It has been suspected for a long time \cite{BFST,FFS,BL,Osborn} 
that in the strongly coupled theories, such as QCD or its 
supersymmetric brethren,  
instantons should be thought of as composite states of more basic
configurations, loosely referred to as `instanton partons'. These partons
would give important and possibly dominant contributions to 
the non-perturbative dynamics at
strong coupling. Our intention is to make this idea more precise.
A necessary evil in our approach is to consider the theory with one of
its dimensions compactified.\footnote{Our approach is different from 
the toron
calculations of Ref.~\cite{CG} where all four dimensions were compactified
on a torus.
The advantage of our method compared to that of \cite{CG} is that we
do not have to fine-tune the compactification parameters for the finite-action 
configurations to exist.
We also note that the value gluino condensate extracted from the 
toron approach of \cite{CG} in the finite-volume torus with the fine-tuned 
periods is difficult to interpret in the infinite volume and its numerical value
agrees neither with the WCI \eqref{stwkb} nor with the SCI 
\eqref{stwka} results. In the alternative toron set-up advocated in 
\cite{Zh}, the fine-tuning problem was avoided at the cost of introducing
{\it singular} toron-like configurations with a branch cut and an IR regulator.} 
To this end, we suppose that, say, 
$x_4$, is periodic with period $\beta/2 \pi$.\footnote{The indices run
over $m=1,2,3,4$ and $\mu=1,2,3$. Our other
conventions in four
and three non-compact dimensions as well as instanton and monopole basics
follow closely Appendices A and C of Ref.~\cite{DKMTV}.} We must then impose 
periodic boundary conditions for bosons and fermions 
\begin{equation}
A_m (x_\mu,x_4=0) \ = \ A_m (x_\mu,x_4=\beta) \ , \quad 
\lambda (x_\mu,x_4=0) \ = \ \lambda (x_\mu,x_4=\beta) \ , 
\elabel{ssbc}\end{equation}
to preserve supersymmetry. An important additional ingredient,
as explained in Sec.~II of \cite{GPY}, is that the local gauge group itself 
must also be composed of `proper' gauge transformations, i.e.~those
that are periodic on $S^1$:
\begin{equation}
U(x_\mu,x_4=0) \ = \ U(x_\mu,x_4=\beta) \ . 
\elabel{ssub}\end{equation}

We will refer to the aforementioned theory as the `theory 
on the cylinder ${\Bbb R}^3 \times S^1$' to distinguish it 
from the finite temperature
compactification where the fermions have {\it anti-periodic\/} 
boundary conditions.
The situation we envisage is similar to that discussed in
\cite{SWthree}, since the theory on
${\Bbb R}^3 \times S^1$ interpolates between the four-dimensional ${\cal N}=1$
pure gauge theory, for $\beta=\infty$, and a three-dimensional
${\cal N}=2$ gauge theory, for $\beta=0$. {}From now on we will work at
finite $\beta$ and only at the end of the calculation
take the limit of $\beta \to \infty$ in order
to recover the genuinely four-dimensional theory.

Some time ago, 
Gross, Pisarski and Yaffe \cite{GPY} gave a complete topological classification of the
smooth finite-action gauge fields on ${\Bbb R}^3 \times S^1$
which may contribute to the path integral in the semi-classical approximation.
The relevant configurations are characterized by three sets of invariants:
the instanton number $k$, the magnetic charge $q$, and the eigenvalues
of the asymptotics of the Wilson line:
\begin{equation}\Omega \ = \ {\Bbb P}\exp \ \int_0^\beta dx_4 \ A_4
(x_4,x_\mu\to \infty)\ .
\elabel{wil}\end{equation}
One consequence of this classification
is that at finite radius $\beta$ instanton 
configurations do not exhaust the set of semi-classical contributions because
configurations with magnetic charges can---and in fact do---contribute to the
non-perturbative dynamics including the value of the gluino condensate.
Ref.~\cite{GPY} further   
argued that in the {\it finite temperature\/} 
compactification---not the one under present consideration---the
non-trivial values of the asymptotic Wilson line \eqref{wil} are suppressed
in the infinite volume limit. Consequently, the classically flat directions 
\begin{equation}
\langle A_4 \rangle = {\rm diag}(a_1,a_2,\ldots,a_N)
\elabel{ppr}\end{equation}
are lifted by thermal
quantum corrections and the true vacuum of the theory is 
$\langle A_4 \rangle =0$.
In this case the configurations with magnetic
charges are not relevant and the semi-classical physics involves 
instantons only. 
Remarkably, for the theory on the cylinder, with periodic boundary
conditions on the fermions, the argument of \cite{GPY} 
does not apply and, as we shall see, the opposite scenario
ensues:    

(i) The semi-classical physics of the 
${\Bbb R}^3 \times S^1$ $SU(N)$ theory is described by configurations of
monopoles of $N$ different types.

(ii) The classical moduli space of the
${\Bbb R}^3 \times S^1$ $SU(N)$ theory \eqref{ppr}
is lifted in a non-trivial way
\begin{equation}
\langle A_4 \rangle \ = \ 
 {\rm diag}\Big({N-1 \over  N}{\pi \over i \beta} \ , 
\  {N-3 \over  N}{\pi \over i \beta} \ ,
\ldots \ , \ 
 -{N-1 \over  N}{\pi \over i \beta}
\Big) \ ,
\elabel{lgen}\end{equation}
leaving behind $N$ supersymmetry-preserving vacua labelled
by the $N$ discrete values of the $\theta$-angle\footnote{In general, the 
$\theta F^{ \ *}F$ term in the microscopic Lagrangian 
can be rotated away with an anomalous chiral transformation of gluinos.
However $\theta$ is an angular variable in the sense that $\theta =
 2\pi n$, $n\in{\Bbb Z}$,
is indistinguishable from $\theta=0$. We will demonstrate in the following that the 
topological charge $Q$ of the configurations contributing to the gluino 
condensate is $Q=1/N$ and thus the net effect of the $\theta F^{ \ *}F$ term
in each vacuum is the phase factor $\theta_u /N$.}
\begin{equation}\theta_u \ = \ 
 2 \pi (u-1) \ , \qquad
u=1,2,\ldots,N\ .
\elabel{ltht}\end{equation}
The $N$ vacua are related to each other by  
the chiral subgroup ${\Bbb Z}_N$, which permutes $\theta_u$'s,
but leaves the Wilson line \eqref{lgen} unchanged.
Each such vacuum
contributes a factor of $1$ to the Witten index
${\rm tr}(-1)^{\rm F}=N$ 
\cite{Windx}.
The values of
gluino condensate in each of these vacua will be related to each other 
by  a trivial phase transformation $\exp[i\theta_u /N]$. 
{}From now on we will concentrate
on simply one of the vacua, with $\theta_u=0$.
The distinctive feature of \eqref{lgen} is the constant equal spacing
between the VEVs $a_j$: 
\begin{equation} a_{j} - a_{j+1} \ = \ {2\pi \over iN\beta} \ {\rm
mod} \ {2\pi\over i\beta}\ ,
\qquad  j=1,2,\ldots,N \ . 
\elabel{lequ}\end{equation}

In general, one would think that
the field configurations of the 
${\Bbb R}^3 \times S^1$ theory which are relevant in the semi-classical regime
are {\it both\/} instantons {\it and\/} monopoles. Remarkably,
however, the instanton configurations are themselves included as
specific multi-monopole configurations. This happens in the following way:
first of all, an instanton configuration
on the cylinder follows from a
standard instanton configuration
in ${\Bbb R}^4$ \cite{BPST} by imposing periodic 
boundary conditions in $x_4$ \eqref{ssbc}. In addition we need
instanton solutions in the presence of a
non-vanishing VEV for the gauge field component
$A_4$, or equivalently a non-trivial expectation of the Wilson line,
as in  Eqs.~\eqref{wil} and \eqref{ppr}. Such
periodic instantons in the presence of a Wilson line were recently analyzed in
Refs.~\cite{LY,KL,LL,KvBzer,KvBone,KvBtwo}.\footnote{For the simpler case of
$\langle A_4 \rangle =0$ periodic instantons were previously constructed 
in Refs.~\cite{GPY,HS}.} It transpires that instantons on the
cylinder can be understood as composite configurations
of $N$ single monopoles, one of each of the $N$ different types
\cite{Nahmtwo,Garland,LY,KL,LL,KvBzer,KvBone,KvBtwo}. One expects in
an $SU(N)$ theory on ${\Bbb R}^4$ that the lowest charged, or 
fundamental, monopoles come in
$N-1$ different varieties, carrying a unit of magnetic charge from
each of the $U(1)$ factors of the $U(1)^{N-1}$ gauge group left
unbroken by the VEV. 
The additional monopole, needed to make up the $N$  types, 
is specific to the compactification on the cylinder
\cite{LY,KL,LL,KvBzer} and will be called here a KK-monpole.
The new monopole carries specific magnetic charges
of the unbroken $U(1)^{N-1}$ gauge group as well as an instanton charge.
The magnetic charge of the KK-monopole is such that
when all $N$ types of monopoles are present, the magnetic charges
cancel and the resulting configuration only carries a unit instanton
charge. 

The $N-1$ fundamental monopoles are the embeddings of the standard
$SU(2)$ BPS monopole \cite{thm,polm,Bog,PS} 
on ${\Bbb R}^3$ spanned by $x_{1,2,3}$ (independent of the $S^1$ coordinate
$x_4$) in the gauge group $SU(N)$. At finite radius  $\beta$, these 
monopoles have finite action and hence contribute 
to the path integral in the semi-classical regime as described in 
Refs.~\cite{Pol,AHW} and \cite{DKMTV,PP,dkmthreed}.
The monopole solutions satisfy Bogomol'nyi equations that
are precisely the 4D self-duality equations\footnote{In the usual
interpretation of the self-duality equations for the monopole, the
time component of the gauge field is interpreted as the Higgs
field; in the present discussion this field {\it is\/} the component
of the gauge field along the compact direction.}
\begin{equation}F_{mn} \ ={ \ }^* F_{mn} \ ,\elabel{seld}\end{equation}
and each solution has two 
adjoint fermion zero-modes as enumerated by the Callias index theorem 
\cite{Cal}. The same consideration applies to the 
KK-monopole
as well, since it is at least formally gauge equivalent to a standard fundamental
monopole via an improper (non-periodic) gauge transformation
\cite{LL}. Since there are two adjoint fermion zero-modes in the
background of each of the $N$ types of monopoles, these configurations
can contribute directly to $\Vev{\tr \lambda^2\over16\pi^2}$.

The remainder of the paper is organized as follows. Initially, we
focus on the case with $SU(2)$ gauge group; the generalization to
$SU(N)$ is then obvious. In general, the
theory on the cylinder can develop a VEV for the
gauge field component along the $S^1$ direction, Eq.~\eqref{ppr}, which for the case
of $SU(2)$ gauge group we parametrize as
\begin{equation}
\langle A_4 \rangle \ = \ v \ {\tau^3 \over 2i} \ \equiv \ 
{\rm diag}\big({v \over 2i} ,-{v \over 2i}\big) \ , \elabel{pptwo}\end{equation}
where $v$ is an arbitrary real parameter
which parametrizes the classical moduli
space. For every fixed $v$ there are actually two distinct vacua
corresponding to the choice of theta angle
Eq.~\eqref{ltht}.
In the Section II, we will show using field theory arguments, backed-up by 
a D-brane analysis, that:

(i) The classical moduli space is a circle,
\begin{equation}
v \in S^1 \ : \quad 0 \le v \le {2 \pi \over \beta} \ . \elabel{clms}\end{equation}
Consequently, $v$ is an angular variable such that 
for any fixed $v\neq 0,2\pi/\beta$, the gauge group is broken to $U(1)$.
 
(ii) There is a conventional `t Hooft-Polyakov BPS monopole 
and an additional `compensating' KK-monopole,
each of which satisfies the self-duality equations \eqref{seld} and 
admits two adjoint fermion 
zero-modes. The singly-charged instanton solution is a composite
configuration of these two monopoles and as expected has
four adjoint fermion zero-modes.

Section III is devoted to an evaluation of the monopole-generated 
superpotential
which has the effect of lifting the classical degeneracy parametrized
by $v$. We argue that the true quantum vacuum state is simply the point
\begin{equation}
v_{\rm vac}= {\pi \over \beta} \ .\elabel{vvac}\end{equation}
Furthermore, at $v=v_{\rm vac}$ the effective potential is zero and 
supersymmetry remains unbroken. 
Hence, there are two supersymmetry-preserving vacua with
\eqref{vvac} and labelled by $\theta_1=0$ and $\theta_2=2\pi$, as per \eqref{ltht},
in agreement with the calculation of the Witten index ${\rm tr}(-1)^{\rm F}=2$ 
\cite{Windx}. 
Moreover, we discover that the superpotential not only lifts the classically
flat direction, but also gives a mass to the dual (magnetic) photon,
which implies confinement of the original electric photon and disappearance
of all the massless modes.

Section IV then goes on to consider the monopole contribution to the gluino condensate.
In the quantum vacua, the gluino condensate $\Vev{\tr
\lambda^2\over16\pi^2}$ receives contributions 
from both the BPS and KK-monopoles. After summing these contributions,
we then take the 
decompactification limit $\beta=\infty$ to obtain the value of 
gluino condensate in the strongly coupled $\N=1$ theory which agrees
with the WCI calculation \eqref{stwkb}. Section V concludes with a
brief discussion.

\rsen
\section{Semi-classical Configurations}

In this section, we consider in more detail the configurations which
contribute to the semi-classical physics for the theory on the
cylinder. We begin with a discussion of the $SU(2)$ case, follow with
an alternative description in terms of D-branes and then indicate how
the results generalize to the $SU(N)$ gauge group.

\subsection{Gauge group $SU(2)$}

To verify that the classical moduli space is $S^1$, consider a
non-periodic (hence `improper') gauge transformation \cite{LL} 
\begin{equation}
U_{\rm special} \ = \ \exp \, \big({\pi x_4 \over i\beta} \tau^3 \big) \ . 
\elabel{lgt}\end{equation}
Improper gauge transformations are treated differently from proper
ones in the path integral, in
the sense that two field configurations related by such a gauge 
transformation do not belong to the same gauge orbit. The transformation
$U_{\rm special}$, however, has a special property: even though it is not itself
periodic,
$U_{\rm special}(x_4=0)=-U_{\rm special}(x_4=\beta)$, the corresponding
gauge transformed field configurations:
\begin{equation}\begin{split}
A'_m \ &= \ \exp \, \big({\pi x_4 \over i\beta} \tau^3 \big) \ 
\left(A_m + \partial_m \right) \ 
\exp \, \big(-{\pi x_4 \over i\beta} \tau^3\big )
 \ , \\
\lambda' \ &= \ \exp \, \big({\pi x_4 \over i\beta} \tau^3\big) \ 
\lambda \ \exp\,  \big (-{\pi x_4 \over i\beta} \tau^3 \big)
 \ , \elabel{sgtc}\end{split}\end{equation}
remain strictly periodic, i.e.
\begin{equation}
A'_m (x_\mu,x_4=0) \ = \ A'_m (x_\mu,x_4=\beta) \ , \quad 
\lambda' (x_\mu,x_4=0) \ = \ \lambda' (x_\mu,x_4=\beta) \ . 
\elabel{prbc}\end{equation}

Applied to the  third component of the gauge field \eqref{pptwo}, 
the transformation $U_{\rm special}$ shifts $v$ according to
\begin{equation}
\langle A'_4 \rangle \ = \ (v - {2\pi \over \beta})
 \ {\tau^3 \over 2i}  \ . \elabel{ppnew}\end{equation}
Thus, using a sequence of these transformation one 
can ratchet-down an arbitrary value of $v \in{\Bbb R}$, to the range
specified in \eqref{clms}. In fact the 
sectors of the theory with $v=\tilde v$ and $v=\tilde v+ 2\pi/\beta$
are physically indistinguishable; one is obtained from the other by relabelling
the Kaluza-Klein modes of the compact direction, i.e.~relabelling
the Matsubara frequencies $\omega_n = 2 n \pi /\beta$ with $n\in{\Bbb Z}$
associated to the compact $x_4\in S^1$ variable.

The standard BPS monopole solution in Hedgehog gauge \cite{Bog,PS} is
\begin{equation}\begin{split}
A^{\sst\rm BPS}_4 (x_\nu)\ &= \ \big(v|x| \ {\rm coth}(v|x|) -1
\big) 
{x_a \over |x|^2}{\tau^a \over 2i} \ , \\
A^{\sst\rm BPS}_\mu (x_\nu)
 \ &= \ \Big(1-{v|x| \over {\rm sinh}(v|x|)}
\Big)\epsilon_{\mu\nu a}{x_\nu \over |x|^2}{\tau^a \over 2i}
 \ .
\elabel{bpscn}\end{split}\end{equation}
These expressions are obviously independent of the 
$S^1$ variable $x_4$,
since the latter can be thought of as the time coordinate of 
the static monopole.
The boundary values of \eqref{bpscn} as $|x| \to \infty$, when gauge
rotated to unitary (singular) gauge, agree with \eqref{pptwo}:
\begin{equation}A^{\sst\rm BPS}_4 \ \to \ v{\tau^3 \over 2i} \ =
\ \langle A_4 \rangle \ . \elabel{bcsi}\end{equation}
The monopole solution \eqref{bpscn} satisfies the self-duality equations
\eqref{seld} and has topological charge
\begin{equation}Q \ \equiv \ {1 \over 16 \pi^2} \ \int_0^\beta dx_4 \int d^3 x
\ \tr\,{}^*F_{mn}F^{mn} \ = \ {\beta v \over 2 \pi} \
. \elabel{topq}\end{equation}
There are precisely two adjoint fermion zero modes 
\cite{Cal} in the monopole background \eqref{bpscn}. 
These modes can be generated by supersymmetry transformations
of the bosonic monopole components in \eqref{bpscn}, yielding
\begin{equation}\lambda^{\sst\rm BPS}_\alpha \ = \ \hf \xi_\beta
(\sigma^m \sigmabar^n)_\alpha^{\ \beta} F^{\sst\rm BPS}_{mn} \ . \elabel{lss}\end{equation}
Here $\sigma^m$ and $\sigmabar^n$ are the four Pauli matrices and 
$\xi_\beta$ is the two-component Grassmann collective coordinate;
see footnote 2.
Finally, the monopole has magnetic charge one,
instanton charge zero, and the action $S_{\sst\rm BPS}$ is
\begin{equation}S_{\sst\rm BPS} \ = \ {4\pi \over g^2} \beta v \ .\elabel{macn}\end{equation}

The monopole of the second type---the KK
monopole---can be obtained \cite{LL} from the expressions \eqref{bpscn} 
by, firstly, replacing the VEV $v$ on the right-hand side of \eqref{bpscn} 
with $2\pi /\beta -v$, and then gauge transforming the resulting expression
with $U_{\rm special}$ as in \eqref{sgtc}. Finally to install the original
VEV $v$ one performs the reflection $v\to -v$ implemented by the discrete
transformation $U_{\rm refl}= \exp [i\pi \tau^2 /4]$.
The resulting configuration is the KK-monopole $A^{\sst\rm KK}_m$
and, though gauge related to $A^{\sst\rm BPS}_m$, it
must, as described earlier, be accounted in the path integral as contributing
to a different topological sector.

The improper gauge transformation $U_{\rm special}$ changes \cite{GPY} 
the instanton charge, $k \to k+q$, and reverses the sign of the magnetic 
charge, $q \to - q$. Thus the KK-monopole
has instanton charge $k=1$ and monopole charge $q=-1$.
The KK-monopole is itself self-dual and its action and topological charge
are:
\begin{equation}\begin{split}S_{\sst\rm KK} \ &= \ {4\pi \over g^2} \beta 
 \ ({2\pi \over \beta} -v) \ , \\
Q_{\sst\rm KK} \ &= \ 1 - {\beta v  \over 2 \pi}
\ .\elabel{makk}\end{split}\end{equation}
As for the original BPS monopole,
there are two adjoint fermion zero-modes (and no anti-fermion
zero-modes) in the KK-monopole background:\footnote{The KK-monopole is
self-dual and not anti-self-dual, and the fact that it has negative
rather than positive magnetic charge is irrelevant for the fermion zero 
mode counting.}
\begin{equation}\lambda^{\sst\rm KK}_\alpha \ = \ \hf \xi_\beta
(\sigma^m \sigmabar^n)_\alpha^{\ \beta} F^{\sst\rm KK}_{mn} \ .
 \elabel{lskk}\end{equation}

As was mentioned earlier, Ref.~\cite{GPY} gave a complete topological 
classification of the
smooth finite-action gauge fields on ${\Bbb R}^3 \times S^1$
in terms of the three invariants:
the instanton number $k$, magnetic charge $q$ and the VEV $v$, in
terms of which
\begin{equation}\begin{split}S_{\rm cl} \ &= \ {8\pi^2 \over g^2} 
 \ (k +q{\beta v \over 2 \pi} ) \ , \\
Q \ &= \ k + q {\beta v  \over 2 \pi}\ .
\elabel{tcln}\end{split}\end{equation}
Comparing Eqs.~\eqref{tcln} for the BPS monopole 
and the KK-monopole with the charges of a single instanton, 
it is tempting to interpret
the latter as the mixed BPS-monopole/KK-monopole configuration.
This interpretation is made precise in Ref.~\cite{KL,LL,KvBzer,KvBone,KvBtwo},
based on earlier work of Refs.~\cite{Nahmtwo,Garland} and \cite{LY}.
We also note that the two gaugino zero-modes of the KK-monopole
combined with the two zero-modes of the BPS-monopole produce the requisite
four adjoint fermion zero-modes of the $SU(2)$ instanton.

\subsection{D-brane description}

Identical conclusions to those reached in Sec.~II.1, can be reached 
in a more geometrical way using D-brane technology and for the
additional insight that this point-of-view provides we describe it
here.\footnote{For a discussion of ${\cal N}=1$ theories and instantons
in the context of branes see Ref.~\cite{Brodie:1998bv}.}

For the geometrical interpretation of the construction, the $\N=4$ case is
most straightforward; the modification
of this set-up relevant to describe the $\N=1$ theory will be considered subsequently.
Therefore we begin with two coincident D3-branes whose collective
dynamics is described by $\N=4$ supersymmetric $SU(2)$ Yang-Mills on
the four-dimensional world volume
\cite{Witpbr,Polch}. 

We now proceed to wrap the world-volume of our D3-branes on the cylinder ${\Bbb R}^3 \times S^1$,
with the radius $R=\beta /2\pi$. With this accomplished, one
performs a T-duality transformation along the compact direction.
The T-duals of the D3-branes are D2-branes stretched along ${\Bbb
R}^3$ and lying orthogonal
to the dual circle $\tilde{S}^1$ with radius $\tilde{R}=\alpha'/R$. 
In the presence of the non-trivial Wilson line Eq.~\eqref{pptwo}, the D2-branes become 
separated by a distance $2\pi \alpha'v$ along the direction of the dual circle \cite{Polch}.
Due to the periodicity around the circle,
we may restrict $0 \le 2\pi \alpha'v \le (2\pi)^2 \alpha' /\beta$,
which is equivalent to Eq.~\eqref{clms}. 

In the T-dual picture, a 
BPS monopole can be represented by 
a D0-brane of length $L_{\sst\rm BPS}=2\pi \alpha' v$ 
stretched between the two D2-branes. The orientation of the D0-brane
(whether the D0-brane is stretched between the first and second
D2-brane, or vice-versa) corresponds
to positive or negative magnetic charge, i.e.~the monopole or anti-monopole.
The monopole mass is the product of the D0-brane tension
$\tau_0=2/(\alpha' g^2)$ and the D0-brane length $L_{\sst\rm BPS}$:
\begin{equation}M_{\sst\rm BPS}\ = \ \tau_0 \ L_{\sst\rm BPS} \ = \ {4 \pi v \over g^2}
 \ ,\ \elabel{mbps}\end{equation}
in agreement with Eq.~\eqref{macn}. 

Actually, as one might have guessed, there is an infinite tower of 
monopoles of the same magnetic
charge coming from the Kaluza-Klein tower over $\tilde{S}^1$ formed by 
wrapping the D0-brane an arbitrary number of times around the
circle. Another way to view the same phenomenon, is to consider the freedom
to add to a monopole a closed
D0-loop starting and ending on the same D2-brane and winding around 
$\tilde{S}^1$; the length of the loop being 
$L_{\rm loop}= (2\pi)^2 \alpha' /\beta $.
This D0-loop over the D2-brane can be identified with the instanton. Indeed, 
after the T-duality transformation along the $\tilde{S}^1$ direction, 
the D2-brane becomes the D3-brane and the D0-loop becomes a
D$(-1)$-brane, or D-instanton.
The D3/D$(-1)$ bound-state is the identified with an instanton having charge equal
the winding number of the D0-loop over $\tilde{S}^1$, and vanishing
magnetic charge. The instanton action is in a similar fashion to Eq.~\eqref{mbps}
\begin{equation}S_{\rm inst} \ = \ \beta M_{\rm inst}\ = \ 
\beta \ \tau_0 \ L_{\rm loop} \ = \ {8 \pi^2 \over g^2}
 \ ,\elabel{sins}\end{equation}
in agreement with Eq.~\eqref{tcln}. In summary, the
standard BPS-monopole is the lowest-lying Kaluza-Klein state with
magnetic charge one: the D0-brane between the first and the second D2-brane.
The monopole of the second type---the KK-monopole---appears as the D0-brane between the second D2-brane
and the first one (hence the magnetic charge $q=-1$) completing the dual circle 
$\tilde{S}^1$ (hence carrying instanton number $k=1$). Furthermore,
the bound-state of the standard and the KK-monopole is the D0-loop,
i.e.~the instanton. Notice that
standard BPS monopole and the KK-monopole (together with their respective
anti-monopoles) are the elemental configurations, out of which the whole
set of semi-classical configurations with arbitrary $k$ and $q$ can be
built.

Although we have described this picture in terms of the $\N=4$ theory,
the whole analysis applies to the $\N=1$ case as well with certain
modifications. 
The $\N=1$ four-dimensional Yang-Mills theory is obtained in a
D-braney fashion from a configuration
of two coincident D4-branes suspended between two NS5-branes, in a
manner described in
Refs.~\cite{EGK,Witqcd}. The world-volume of the D4-branes is infinite
in four directions ${\Bbb R}^4$ and is finite in the fifth direction
$\Delta_5$, which is
the separation between the NS5-branes along the D4-branes.
Following  the same line of reasoning  as in the $\N=4$ case,
the {\it infinite\/} part of the world-volume of the D4-branes is put on the
cylinder and T-dualized. The T-duals of the D4-branes on 
$\Delta_5\times {\Bbb R}^3 \times S^1$
are the D3-branes stretched along $\Delta_5\times {\Bbb R}^3$ and orthogonal
to the dual circle $\tilde S^1$ with the dual radius $\tilde{R}=\alpha'/R$. 
In the
presence of the non-trivial Wilson line Eq.~\eqref{pptwo}, the D3-branes become 
separated \cite{Polch} by the distance $2\pi \alpha'v$ along the dual circle,
which is equivalent to Eq.~\eqref{clms}. 

In the $\N=4$ theory the D2-branes are BPS configurations and
consequently, when at rest, there is no interaction
between them. Thus, their separation along the dual circle is
arbitrary, i.e.~$v$ is an arbitrary modulus.
In the $\N=1$ theory, the D3-branes are {\it not\/} BPS configurations.
In the next Section, via an explicit calculation of a
superpotential, we will prove that they
actually repel each other. Geometrically this implies that the two D3-branes
stay at the opposite ends of the dual circle and consequently $v=\pi/\beta$.
Hence, the classically flat direction is lifted precisely in the
manner predicted by Eq.~\eqref{vvac}.

The previous set-up described in the context of $SU(2)$ can be immediately
generalized to $SU(N)$.
We now have $N$ D3-branes positioned along a circle and repelling
each other; hence one expects 
\begin{equation} a_{j} - a_{j+1} \ = \ {2\pi \over iN\beta} \ {\rm
mod} \ {2\pi\over i\beta}\ ,
\qquad j=1,2,\ldots,N \ , 
\elabel{lenw}\end{equation}
and hence, \eqref{lgen}. The $N$ types of the monopole-like
configurations are composed from $N-1$ standard BPS monopoles represented
by the D1-branes of minimal lengths stretched between the adjacent
$u^{\rm th}$ and $(u+1)^{\rm th}$ D3-branes, $u=1,\ldots,N-1$. The
KK-monopole is the D1-brane stretched
between the $N^{\rm th}$ and $1^{\rm st}$ D3-branes. The instanton, as
before, is the closed D1-loop around the $\tilde{S}^1$ direction.

\rsen
\section{Evaluation of the Superpotential}

In this section, we will determine the superpotential of the $\N=1$ 
supersymmetric
$SU(2)$ Yang-Mills theory on ${\Bbb R}^3 \times S^1$. The superpotential 
is trivial
in perturbation theory, but receives non-perturbative contributions
as described in Ref.~\cite{AHW}. Contributions arise from both types
of monopole: BPS and KK. As advertised earlier, the classical moduli
space \eqref{clms} will be lifted by
the superpotential in accordance with Eqs.~\eqref{lgen} and \eqref{vvac}.

In the presence of a non-vanishing VEV $v$, fields with isospin components not
aligned with the scalar VEV Eq.~\eqref{pptwo}, acquire masses $m \propto v$
via the Higgs
mechanism. The massless fields are consequently the $U(1)$ projections 
$A_m^3$ and $\lambda^3$ of the non-abelian fields $A_m^a$ and
$\lambda^a$, $a=1,2,3$.
Moreover, since $x_4$ is periodic, each field can
be Fourier analyzed as an expansion over the discrete Matsubara frequencies 
$\omega_n=2\pi n/\beta$ with all the $n \neq 0$ modes being massive
Kaluza-Klein modes. The $n=0$ modes correspond to the fields independent
of the $x_4$ coordinate. Thus, the classically massless degrees-of-freedom
are the $x_4$-independent $U(1)$ fields $A_\mu (x_\nu)$, $\phi(x_\nu)$, 
$\chi(x_\nu)$ and $\chitilde(x_\nu)$ defined\footnote{Note that
the calculations in Appendix A of Ref.~\cite{DKMTV} and the details of the
comactification to 3D were given in Minkowski space with $x_3$ and not $x_4$
being the compactified direction. Here we analytically continue the results
of \cite{DKMTV} to Euclidean space.} as in Appendix A
of Ref.~\cite{DKMTV}:
\begin{subequations}\begin{align}
A_\mu \ &= \ A_\mu^3 \ : \quad {\rm for} \ \mu=1,2,3 \ , \elabel{adef}\\
\phi \ &= \ A_4^3 \ , \elabel{phid}\\
\chi_\alpha \ &= \ {1 \over \sqrt{2}}(\lambda^3_\alpha
+\lambdabar^3_{\dalpha}) \ , \qquad
\chitilde_\alpha \ = \ -{i \over \sqrt{2}}(\lambda^3_\alpha
-\lambdabar^3_{\dalpha}) \ .
\elabel{chde}\end{align}\end{subequations}
Here $\chi$ and $\chitilde$ are the Majorana two-spinors in three-dimensions.
The classical action for the massless fields $S_{\rm cl}^{U(1)}$
can be read-off from the four-dimensional
action of $\N=1$ supersymmetric Yang-Mills (cf.~\cite{DKMTV}):
\begin{equation}S_{\rm cl}^{U(1)} \ = \ {\beta \over g^2} \int d^3 x
\big(\tfrac14 F_{\mu \nu} F^{\mu \nu} + \tfrac12 \partial_\mu \phi
\partial^\mu \phi - \tfrac12 \chi \hat{\partial} \chi
- \tfrac12 \chitilde \hat{\partial} \chitilde \big)
 \ , \elabel{smls}\end{equation}
where $\hat{\partial}=\gamma^\mu\partial_\mu$, and $\gamma^\mu$ are
the three-dimensional gamma-matrices.
The presence of the monopoles in the microscopic theory means we must also
include a surface term $S_{\rm sf}$ in the action \eqref{smls}: 
\begin{equation}S_{\rm sf} \ = \ -{i\sigma \beta \over 8 \pi} \int d^3x 
 \ \epsilon^{\mu\nu\rho}\partial_\mu F_{\nu \rho} \ .\elabel{ssur}\end{equation}
Due to Dirac quantization of magnetic charge:
\begin{equation}
q \ = \ {1 \over 8 \pi} \int d^3x 
 \ \epsilon^{\mu\nu\rho}\partial_\mu F_{\nu \rho} \ \in \ {\Bbb Z} \ ,
\elabel{qdef}\end{equation}
in \eqref{ssur} $\sigma$ is a periodic Lagrange multiplier variable with period 
$2\pi/\beta$. 

Following Polyakov \cite{Pol} an equivalent dual description of the low-energy
theory \eqref{smls} and \eqref{ssur} can be obtained by promoting $\sigma$ to be a dynamical
field $\sigma(x)$. This field serves as the Lagrange multiplier for the
Bianchi identity constraint. The classical action for massless fields
then contains the terms
\begin{equation}\beta \int d^3 x
\big({1 \over 4g^2} F_{\mu \nu} F^{\mu \nu} -i{\sigma \over 8 \pi}
\epsilon^{\mu\nu\rho}\partial_\mu F_{\nu \rho} + \cdots \big) \ .\elabel{phts}\end{equation}
At this stage the photon field-strength $F_{\mu\nu}(x)$ can be integrated out,
and the resulting classical massless action reads
\begin{equation}S_{\rm cl} \ = \ {\beta \over g^2} \int d^3 x
\big(\tfrac12\partial_\mu \gamma \partial^\mu \gamma 
+ \tfrac12 \partial_\mu \phi
\partial^\mu \phi - \tfrac12 \chi \hat{\partial} \chi
-  \tfrac12 \chitilde \hat{\partial}\chitilde \big)
 \ , \elabel{scmp}\end{equation}
where we have introduced the dual photon scalar field $\gamma(x)$:
\begin{equation}\gamma(x)\ = \ {g^2 \over 4\pi} \ \sigma(x) \ .\elabel{dphd}\end{equation}
This action is invariant under infinitesimal $\N=2$ supersymmetry transformations in three dimensions:
\begin{equation}\begin{split}\delta \phi \ &= \ \sqrt{2}\xi_1^\alpha \chi_\alpha
- \sqrt{2}\xi_2^\alpha \chitilde_\alpha \ , \\
\delta \gamma \ &= \ \sqrt{2}\xi_1^\alpha \chitilde_\alpha
+ \sqrt{2}\xi_2^\alpha \chi_\alpha \ , \\
\delta \chi^\alpha \ &= \ \sqrt{2}\xi_1^\beta \hat{\partial}_\beta^{\ \alpha}
\phi + \sqrt{2}\xi_2^\beta \hat{\partial}_\beta^{\ \alpha}\gamma \ , \\
\delta \chitilde^\alpha \ &= \ \sqrt{2}\xi_1^\beta 
\hat{\partial}_\beta^{\ \alpha}\gamma 
- \sqrt{2}\xi_2^\beta \hat{\partial}_\beta^{\ \alpha}\phi \ .
\elabel{sctr}\end{split}\end{equation}

It is more convenient for our purposes to use a more compact form 
of Eqs.~\eqref{scmp} and \eqref{sctr} involving the complex
complex scalar $Z$ and fermion $\Psi$:
\begin{equation}\begin{split}
Z \ = \ \phi + i \gamma \ , \qquad 
&\bar{Z} \ = \ \phi - i \gamma \ , \\
\Psi_\alpha \ = \ \chi_\alpha + i \chitilde_\alpha \ , \qquad
&\bar{\Psi}_\alpha \ = \ \chi_\alpha - i \chitilde_\alpha \ , 
\elabel{zps}\end{split}\end{equation}
in terms of which the action is
\begin{equation}S_{\rm cl} \ = \ {\beta \over g^2} \int d^3 x
\big( \tfrac12\partial_\mu \bar{Z}
\partial^\mu Z - \tfrac12\bar{\Psi}  \hat{\partial}\Psi \big) \ , 
\elabel{scmm}\end{equation}
and the supersymmetry transformations are
\begin{equation}\begin{split}
\delta Z \ = \ \sqrt{2}\theta^\alpha \Psi_\alpha
 \ , \qquad 
&\delta\bar{Z} \ = \ \sqrt{2}\bar{\theta}^\alpha \bar{\Psi}_\alpha
 \ , \\
\delta\Psi^\alpha \ = \ \sqrt{2}(\bar{\theta}\hat{\partial})^\alpha Z
 \ , \qquad
&\delta\bar{\Psi}^\alpha \ = \ \sqrt{2}(\theta \hat{\partial})^\alpha \bar{Z}
 \ ,\elabel{stzp}\end{split}\end{equation}
where we have introduced the infinitesimal supersymmetry transformation parameter
$\theta^\alpha=\xi_1^\alpha +i\xi_2^\alpha$.

Non-perturbative quantum effects will modify the classical action
for massless fields, Eq.~\eqref{scmm}, by generating a superpotential 
${\cal W}(\Phi)$ and ${\bar{\cal W}}(\bar{\Phi})$ written in terms of the 
chiral and anti-chiral $\N=1$ superfields:
\begin{equation}\begin{split}\Phi \ &= \ Z+\sqrt{2}\theta^\alpha \Psi_\alpha
+ \theta^\alpha \theta_\alpha {\cal F} \ , \\
\bar{\Phi} \ &= \ \bar{Z}+\sqrt{2}\bar{\theta}^\alpha \bar{\Psi}_\alpha
+ \bar{\theta}^\alpha \bar{\theta}_\alpha \bar{\cal F} \ .
\elabel{sfde}\end{split}\end{equation}
With the addition of this superpotential, the 
resulting quantum low-energy effective action reads
\begin{equation}S_{\rm eff} \ = \ S_{\rm cl} \ + \ 
{\beta \over g^2} \int d^3 x\, \Big(\int d^2 \theta\, {\cal W}(\Phi)
\ + \ \int d^2 \bar{\theta}\, {\bar{\cal W}}(\bar{\Phi}) \Big) \ . \elabel{sptg}\end{equation}
As usual, the scalar potential $V_{\rm eff}$ is determined by the
derivatives of the superpotential
with respect to the scalar fields
\begin{equation}V_{\rm eff} \ = \ {\cal F}\bar{\cal F} \ = \ 
{\partial {\cal W} \over \partial Z}
{\partial \bar{\cal W} \over \partial\bar{Z}} \ .\elabel{vefd}\end{equation}
The true vacuum corresponds to the minimum of $V_{\rm eff}$. In general
$V_{\rm eff}\ge 0$, and supersymmetry is unbroken only if the vacuum
solution has $V_{\rm eff}= 0$.

We are now ready to calculate the superpotential ${\cal W}(\Phi)$
and hence the true ground state of the theory in the semi-classical approximation.
Since the standard BPS monopole and the KK-monopole have two fermion 
zero-modes apiece, Eqs.~\eqref{lss} and \eqref{lskk}, they both 
generate mass terms
for classically massless fermions $\bar{\Psi}$, while the corresponding
anti-monopoles will generate a mass for $\Psi$:
\begin{equation}{\cal L}_{\rm mass} \ = \ 
{m_{\bar{\Psi}} \over 2} \bar{\Psi}\bar{\Psi} \ + \ 
{m_{\Psi} \over 2} \Psi\Psi \ = \ 
m_{\bar{\Psi}}  \lambdabar^3\lambdabar^3 \ + \ 
m_{\Psi}  \lambda^3\lambda^3 \ .\elabel{sems}\end{equation}
The supersymmetric completion
of \eqref{sems} in the low-energy effective action will give 
the superpotential in question.
The masses \eqref{sems} are determined by examining the large distance behaviour 
of the correlators
\begin{subequations}\begin{align}
G^{(2)}_{\alpha\beta}(x,y) \ &= \ 
\langle\lambda^3_\alpha(x)\lambda^3_\beta(y)\rangle \ ,\elabel{lcor}\\
\bar{G}^{(2)}_{\alpha\beta}(x,y) \ &= \ 
\langle\lambdabar^3_\alpha(x)\lambdabar^3_\beta(y)\rangle \
. \elabel{lbco}
\end{align}\end{subequations}
Using the LSZ reduction formulae, somewhat along the lines of
Ref.~\cite{AHW}, we find
\begin{subequations}\begin{align}
G^{(2)}_{\alpha\beta}(x,y) \ &\to \ 2m_{\bar{\Psi}} \ 
\beta\int d^3 X \, {\cal S}_{\rm F}(x-X)_{\alpha \rho}\epsilon^{\rho \delta}
{\cal S}_{\rm F}(y-X)_{\beta \delta} \ , \elabel{mbar}\\
\bar{G}^{(2)}_{\alpha\beta}(x,y) \ &\to \ 2m_{\Psi} \ 
\beta\int d^3 X \, {\cal S}_{\rm F}(x-X)_{\alpha \rho}\epsilon^{\rho \delta}
{\cal S}_{\rm F}(y-X)_{\beta \delta} \ . \elabel{mmm}
\end{align}\end{subequations}
Here ${\cal S}_{\rm F}(x)$ is the massless fermion propagator in 3D, or,
equivalently, the Weyl-fermion propagator on ${\Bbb R}^3\times S^1$ with 
zero Matsubara frequency: ${\cal S}_{\rm F}(x)=\gamma^\mu x_\mu/(4\pi|x|)^2$.

We first consider the contribution of a single standard BPS-monopole,
\eqref{bpscn} and \eqref{lss}, to $m_{\bar{\Psi}}$:
\begin{equation}\langle\lambda^3_\alpha(x)\lambda^3_\beta(y)\rangle_{\sst\rm BPS}
 \ = \ 
\int d\mu^{\sst\rm BPS} \lambda^{\sst\rm LD}_\alpha(x)\lambda^{\sst\rm LD}_\beta(y)
\ ,\elabel{lccr}\end{equation}
where $\lambda^{\sst\rm LD}_\alpha(x)$ is the large distance (LD) limit
of the fermion zero modes \eqref{lss} as computed in Appendix C of Ref.~\cite{DKMTV}: 
\begin{equation}\lambda^{\sst\rm LD}_\alpha(x)\ = \ 
8\pi {\cal S}_{\rm F}(x-X)_{\alpha}^{\ \rho}\xi_\rho \ ,\elabel{fzld}\end{equation}
and $d\mu^{\sst\rm BPS}$ is the semiclassical integration measure
of the standard single-monopole on ${\Bbb R}^3\times S^1$:
\begin{equation}
\int d\mu^{\sst\rm BPS} \ = \ M_{\sst\rm PV}^3 \ e^{-S_{\sst\rm BPS}} \
\int {d^3 X \over (2\pi)^{3/2}}[g^2 S_{\sst\rm BPS}]^{3/2} \ 
\int_0^{2\pi} {d \Omega \over \sqrt{2\pi}}[g^2 S_{\sst\rm BPS}/v^2]^{1/2} \
\int d^2 \xi {1 \over 2g^2 S_{\sst\rm BPS}} \ .\elabel{msst}\end{equation}
This measure is obtained in the standard way by changing variables
in the path integral from field fluctuations around the monopole
to the monopole's collective coordinates: $X_{\mu}$ (position),
$\Omega$ ($U(1)$-angle) and $\xi_\alpha$ (Grassmann collective coordinates).
The relevant Jacobian factors in \eqref{msst} are taken from Ref.~\cite{DKMTV}.
In contradistinction with the 3D calculation of \cite{DKMTV}, our
present calculation is locally four-dimensional, i.e~in the path integral
we have integrated over the fluctuations around the monopole configuration
in ${\Bbb R}^3\times S^1$. Thus, the UV-regularized
determinants over non-zero eigenvalues of the
quadratic fluctuation operators cancel between fermions and bosons
due to supersymmetry as in Ref.~\cite{Adda}.\footnote{In order to invoke the
result of \cite{Adda} one needs the self-duality of the solution, a covariant background gauge,
four dimensions and supersymmetry.}
The ultra-violet divergences are regularized in the Pauli-Villars scheme,
which explains the appearance of the Pauli-Villars mass scale
$M_{\sst\rm PV}$ to a power given by $n_{\rm b} -n_{\rm f}/2=3$, where 
$n_{\rm b}=4$ and $n_{\rm f}=2$ are, respectively, the numbers of 
bosonic and fermionic zero-modes of the monopole.
Collecting together the expressions in 
Eqs.~\eqref{macn}, \eqref{mbar}, \eqref{lccr}, \eqref{fzld} and
\eqref{msst}, we find the single-monopole
contribution to $m_{\bar{\Psi}}$ is 
\begin{equation}m_{\bar{\Psi}}^{\sst\rm BPS} \ = \ 16 \pi^2 \beta^2 M_{\sst\rm PV}^3
\exp\big[-{4\pi \over g^2} \beta v\big] \ .\elabel{ssmm}\end{equation}

This expression ignores the contributions of 
monopole--anti-monopole pairs in the 
background of the single monopole configuration and since 
monopole--anti-monopole interactions are long-range (Coulombic) their
effects are considerable and must be taken into account. This is
precisely the famous Polyakov effect \cite{Pol} and fortunately
there is a very elegant way to incorporate it. The interactions of a single
monopole with the monopole--anti-monopole medium can be taken into
account in a way by simply coupling the monopole to the magnetic photon
$\gamma(x)$ (or $\sigma(x)$) introduced earlier, in Eqs.~\eqref{ssur} and \eqref{dphd}, 
and at the same time promoting the VEV $v$ to a dynamical scalar field 
$\phi(x)$. The coupling of the dual photon to the monopole of magnetic
charge $q$ is dictated by the surface term in Eq.~\eqref{ssur}. Naturally enough, one
is instructed \cite{AHW,Pol} to change
the action \eqref{tcln} of the original semi-classical configuration as follows:
\begin{equation}
S_{\rm cl} \ = \ {8\pi^2 \over g^2} 
 \ \big(k +q{\beta v \over 2 \pi} \big) \ \rightarrow \ 
{8\pi^2 \over g^2} 
 \ \big(k +q{\beta \phi(x) \over 2 \pi} \big) \ + \ 
i q\beta \sigma(x) \ . \elabel{nact}\end{equation}
This means that the mass becomes a local coupling:
\begin{equation}m_{\bar{\Psi}}^{\sst\rm BPS}(x) \ = \ 16 \pi^2 \beta^2 M_{\sst\rm PV}^3
\exp\big[-{4\pi\over g^2}\beta \phi(x) + 
i{4\pi\over g^2}\beta \gamma(x)\big] \ .\elabel{locc}\end{equation}

It is straightforward to derive 
the single KK-monopole contribution
to $m_{\bar{\Psi}}$. It is obtained in the same way as the expression 
on the right hand side of \eqref{ssmm}, but instead of $S_{\sst\rm BPS}$ in \eqref{msst} 
one has to use $S_{\sst\rm KK}$ of \eqref{makk}:\footnote{This is because the KK-monopole is gauge equivalent
to the standard monopole with a `wrong' VEV as explained in the Sec.~II.1}
\begin{equation}
m_{\bar{\Psi}}^{\sst\rm KK} \ = \ 16 \pi^2 \beta^2 M_{\sst\rm PV}^3
\exp\big[-{4\pi \over g^2} (2\pi -\beta v)\big] \ .\elabel{sskk}\end{equation}

The total mass coupling $m_{\bar{\Psi}}(x)$ is then given by the sum of the 
standard BPS- and the KK-monopole contributions, each embedded into
the dual magnetic field theory as per Eq.~\eqref{nact}: 
\begin{equation}\begin{split}
m_{\bar{\Psi}}(x) \ = &\ 16 \pi^2 \beta^2 M_{\sst\rm PV}^3 \\ 
&\times\Big(\exp\big[-{4\pi\over g^2}\beta \phi(x) + 
i{4\pi\over g^2}\beta \gamma(x)\big] \ + \ 
\exp\big[-{8\pi^2 \over g^2}+{4\pi\over g^2}\beta \phi(x) - 
i{4\pi\over g^2}\beta \gamma(x)\big]
\Big) \ .\elabel{mtot}\end{split}\end{equation}
In the second term above, we used the fact that the KK-monopole has 
$q_{\sst\rm KK}=-1$ and $k_{\sst\rm KK}=1$.
Denoting the overall coefficient in \eqref{mtot} as $M$:
\begin{equation}M \ \equiv \ 16 \pi^2 \beta^2 M_{\sst\rm PV}^3\
,\elabel{mdfn}
\end{equation}
and making use of the complex scalar field and fermion of \eqref{zps} we finally 
get the following expression for the mass term:
\begin{equation}{\cal L}_{\rm mass} \ = \ 
{M \over 2} \ \bar{\Psi}(x)\bar{\Psi}(x) \ \Big(
\exp\big[-{4\pi\beta\over g^2}\bar{Z}(x)\big] \ + \ 
\exp\big[-{8\pi^2\over g^2}+{4\pi\beta\over g^2}\bar{Z}(x)\big]
\Big) \ .\elabel{mtps}\end{equation}
This coupling corresponds to
a superpotential ${\bar{\cal W}}(\bar{\Phi})$ term in \eqref{sptg} 
of the form:
\begin{equation}{\bar{\cal W}}(\bar{\Phi}) \ = \ 
\Big({g^2 \over 4\pi\beta} \Big)^2 \ M \ \Big(
\exp\big[-{4\pi\beta\over g^2}\bar{\Phi}\big] \ + \ 
\exp\big[-{8\pi^2\over g^2}+{4\pi\beta\over g^2}\bar{\Phi}\big]
\Big) \ .\elabel{spdf}\end{equation}
Equivalently, the anti-monopoles generate the hermitian conjugate:
\begin{equation}{\cal W}(\Phi) \ = \ 
\Big({g^2 \over 4\pi\beta} \Big)^2 \ M \ \Big(
\exp\big[-{4\pi\beta\over g^2}\Phi\big] \ + \ 
\exp\big[-{8\pi^2\over g^2}+{4\pi\beta\over g^2}\Phi\big]
\Big) \ .\elabel{spdh}\end{equation}

With the expression for the superpotential in hand,
we can now calculate the scalar potential \eqref{vefd} and determine 
the true vacuum state of the theory. Consider
\begin{equation}{\cal F} \ = \ {\partial {\cal W} \over \partial Z}
\ =  \ -{ M g^2 \over 4\pi\beta} \  \Big(
\exp\big[-{4\pi\beta\over g^2} Z\big] \ - \ 
\exp\big[-{8\pi^2\over g^2}+{4\pi\beta\over g^2} Z\big]
\Big) \ .\elabel{fded}\end{equation}
The supersymmetry preserving vacuum 
$\langle Z \rangle=\langle \phi \rangle+i\langle \gamma \rangle$
corresponds to 
\begin{equation}
{\cal F}(\langle Z \rangle) \ = \ 0  \ \quad \implies\quad 
\langle Z \rangle \ = \ {\pi \over \beta} \ , \elabel{spvc}\end{equation}
which corresponds to the scalar VEV 
$\langle \phi \rangle\equiv v=\pi/\beta$ as predicted in \eqref{vvac}.

Note that since $\langle \gamma \rangle=0$ the dual photon 
does not condense, as expected. 
What is much more interesting is that the dual photon becomes massive, 
\begin{equation}V_{\rm eff} (\phi={\pi \over \beta}, \gamma(x)) \ = \ 
2  \ \Big({M g^2 \over 4\pi \beta }\Big)^2 
\exp\big[-{8\pi^2 \over g^2}\big] \ 
\big(1 - \cos {8\pi\beta \over g^2}\gamma(x) \big) \ , \elabel{vefg}\end{equation}
which implies confinement of the original electric photon
and the corresponding disappearance of all the massless modes.

\rsen
\section{Gluino Condensate from Monopoles}

In this section, we use our description of the quantum vacuum state of
the theory to evaluate the monopole contribution to the gluino condensate.

\subsection{Gauge group $SU(2)$}

We are now in a position to directly compute gluino condensate in
the $SU(2)$ theory.
Firstly, we evaluate the standard BPS monopole contribution 
to $\Vev{\tr \lambda^2}$:
\begin{equation}\begin{split}
\VEV{\tr \lambda^2}_{\sst\rm BPS} 
=\ M_{\sst\rm PV}^3 \ e^{-S_{\sst\rm BPS}} \
&\int {d^3 X \over (2\pi)^{3/2}}[g^2S_{\sst\rm BPS}]^{3/2} \ 
\int_0^{2\pi} {d \Omega \over \sqrt{2\pi}}[g^2S_{\sst\rm BPS}/v^2]^{1/2} \\ 
&\times\int d^2 \xi {1 \over 2g^2S_{\sst\rm BPS}} \ 
 {\rm tr}( \lambda^{{\sst\rm BPS} \ \alpha}(x) 
\lambda^{\sst\rm BPS}_\alpha(x) ) \ ,
\elabel{glbp}\end{split}\end{equation}
where we have used the expression \eqref{msst} for the monopole measure.
To evaluate \eqref{glbp},  we use the normalization of fermion zero modes from
Ref.~\cite{DKMTV}:
\begin{equation}\int d^3 X \int d^2 \xi  \ 
{\rm tr}\left( \lambda^{{\sst\rm BPS} \ \alpha}(x) 
\lambda^{\sst\rm BPS}_\alpha(x) \right) \ = \ 2S_{\sst\rm BPS} \ {g^2 \over \beta} \ .
\elabel{nfzm}\end{equation}
A straightforward calculation gives
\begin{equation}
\VEV{\tr \lambda^2\over16\pi^2}_{\sst\rm BPS} \ = \ {1 \over 2} {\beta v \over \pi}
M_{\sst\rm PV}^3 \exp\big[-{8\pi^2 \over g^2}{\beta v \over 2\pi}\big] \ . \elabel{lcst}\end{equation}
The KK-monopole contribution is obtained by changing
$S_{\sst\rm BPS} \rightarrow S_{\sst\rm KK}$ in the BPS-expressions
above to give
\begin{equation}
\VEV{\tr \lambda^2\over16\pi^2}_{\sst\rm KK} \ = \ {1 \over 2} \ \big(2- {\beta v \over \pi}\big)
 \ M_{\sst\rm PV}^3 \exp\big[-{8\pi^2 \over g^2}
+{8\pi^2 \over g^2}{\beta v \over 2\pi}\big] \ . \elabel{llkk}\end{equation}

The expressions \eqref{lcst} and \eqref{llkk} explicitly depend on the UV-cutoff
$M_{\sst\rm PV}$ and do not appear to be renormalization group invariant.
However, it is pleasing that in the true ground-state established in
the last section, this worrisome dependence disappears. At 
$v=\pi\beta$, we get
\begin{equation}
\VEV{\tr \lambda^2\over16\pi^2}_{\sst\rm BPS} \ =\ 
\VEV{\tr \lambda^2\over16\pi^2}_{\sst\rm KK}\ =\ 
\hf M_{\sst\rm PV}^3 \exp\big[-{4\pi^2 \over g^2}\big] \ .\elabel{lctw}\end{equation}
Finally, introducing the renormalization group invariant scale
$\Lambda_{\sst\rm PV}$ of the theory via
\begin{equation}
M_{\sst\rm PV}^3 \exp\big[-{4\pi^2 \over g^2}\big]
\ = \ \Lambda^3 \ , \elabel{lamd}\end{equation}
and adding together both monopole contributions we obtain a
value for the gluino condensate:
\begin{equation}\VEV{\tr \lambda^2\over16\pi^2}\ =\ \Lambda^3  \ .
\elabel{restw}\end{equation}
This is precisely the value obtained in the WCI approach \eqref{stwkb}.

\subsection{Generalization to $SU(N)$}

The calculation of the superpotential and the gluino condensate can be
straightforwardly generalized to the 
case of $SU(N)$ gauge group. The quantum vacuum has
\begin{equation} a_{j} - a_{j+1} \ = \ {2\pi \over iN\beta} \ {\rm
mod} \ {2\pi\over i\beta}\ ,
\qquad j=1,2,\ldots,N \ , 
\elabel{lnnn}\end{equation}
and so each of the $N$ types of monopoles ($N-1$ standard BPS and one
KK) have equal actions and equal topological charges:
\begin{equation}
S_{\rm mono} \ = \ {8 \pi^2 \over N g^2} \ , \qquad
Q_{\rm mono} \ = \ {1\over N }
 \ .\elabel{smon}\end{equation}
The contribution of a single monopole to the gluino condensate will be,
in analogy with Eq.~\eqref{lctw}, 
\begin{equation}
\VEV{\tr \lambda^2\over16\pi^2}_{\sst\rm BPS}  \ = \ 
\VEV{\tr \lambda^2\over16\pi^2}_{\sst\rm KK}\ =
\ {1 \over N} 
M_{\sst\rm PV}^3 \exp\big[-{8\pi^2 \over N g^2}\big] \ . \elabel{lcth}\end{equation}
The first coefficient of the $\beta$-function is now
$b_0=3N$ and the analog of \eqref{lamd} reads
\begin{equation}
M_{\sst\rm PV}^{3N} \exp\big[-{8\pi^2 \over g^2}\big]
\ = \ \Lambda^{3N} \ . \elabel{latwo}\end{equation}
Finally, the total contribution of the $N$ monopoles to the gluino
condensate, as in $SU(2)$, reproduces the WCI value \eqref{stwkb}.

\rsen
\section{Discussion}



More than twenty years ago Polyakov \cite{Pol} famously observed that in 
three-dimensional gauge-Higgs theory without fermions 
the magnetic 
photon $\gamma(x)$ gets a non-zero mass due to 
Debye screening in the monopole--anti-monopole plasma. The mass term 
for the dual photon then implies confinement of the original electric 
photon. 
A na\"\i ve attempt to generalize this mechanism to four 
dimensions by simply substituting the three-dimensional instantons 
(i.e.~monopoles) with four-dimensional instantons fails,
since in four dimensions instantons and anti-instantons
have a dipole--dipole interaction which is short-ranged
and hence, the instanton--anti-instanton medium cannot
form a Coulomb plasma essential for Polyakov's Debye mechanism.
However, it has been suspected for a long time that instantons
and anti-instantons can be thought of as composite states of more basic 
configurations---instanton partons---which would 
have long-range interactions and lead to a Coulomb plasma and to the Debye screening.

In this paper, following earlier ideas of \cite{LY,KL,LL,KvBzer,KvBone,KvBtwo},
we have identified the instanton partons with monopoles in the
four-dimensional gauge theory compactified on ${\Bbb R}^3 \times S^1$.
The Debye screening in the monopole plasma induces a non-zero
mass for the dual photon.
Hence, we have successfully generalized Polyakov's mechanism of confinement
to the four-dimensional supersymmetric gauge theory compactified
on ${\Bbb R}^3 \times S^1$.

As the VEVs in Eq.~\eqref{lgen}
are inversely proportional to the radius $\beta$,
the theory becomes weakly coupled at small $\beta$ and can be analysed
semi-classically. To return to the strongly coupled theory in
Minkowski space, we need to consider the opposite limit of large $\beta$.
Since all the F-terms are holomorphic functions of the fields and since
the VEVs of the fields \eqref{lgen} are holomorphic functions of $\beta$,
the power of holomorphy \cite{Seiberg93} allows to analytically continue
the semi-classical values of the F-terms to the strong-coupling regime. 
 
As a useful practical application and a test of monopole physics in 
${\Bbb R}^3 \times S^1$, we have calculated the value of gluino condensate
and taken the decompactification limit to reproduce the WCI result.

\medskip

\centerline{$\sst**********************$}

\medskip

We thank Diego Bellisai, Pierre van Baal and Misha Shifman
 for valuable discussions.
VVK and MPM acknowledge a NATO Collaborative Research Grant,
TJH and VVK acknowledge the TMR network grant FMRX-CT96-0012
and NMD acknowledges a PPARC studentship.


\begin{thebibliography}{99}
\bibitem{HKLM}{T.J. Hollowood, V.V. Khoze, W. Lee and M.P. Mattis,
``Breakdown of Cluster Decomposition in Instanton Calculations of the Gaugino
Condensate'', {\tt hep-th/9904116}.}

\bibitem{LY}{K. Lee and P. Yi, 
Phys. Rev. {\bf D56} (1997) 
3711, {\tt hep-th/9702107}.}

\bibitem{KL}{K. Lee, 
Phys. Lett. {\bf B426} (1988) 323,
    {\tt hep-th/9802012}.}

\bibitem{LL}{
 K. Lee and C. Lu, 
 Phys. Rev. {\bf D58} (1988) 025011,
    {\tt hep-th/9802108}.}

\bibitem{KvBzer}{T.C. Kraan and P. van Baal, 
 Phys. Lett. {\bf B428} (1998) 268, {\tt hep-th/9802049}.}

\bibitem{KvBone}{T.C. Kraan and P. van Baal, 
 Nucl. Phys. {\bf B533} (1998) 627, {\tt hep-th/9805168}.}

\bibitem{KvBtwo}{T.C. Kraan and P. van Baal, 
Phys. Lett. {\bf B435} (1998) 389,  {\tt hep-th/9806034}.} 



\bibitem{NSVZone}{V.A. Novikov, M.A. Shifman, A.I. Vainshtein and
V.I. Zakharov, Nucl. Phys. {\bf B229} (1983) 394; 
Nucl. Phys. {\bf B229} (1983) 407.}

\bibitem{ARV}{G.C. Rossi and G. Veneziano, Phys. Lett. {\bf 138B} 195;
D. Amati, G.C. Rossi and G. Veneziano, 
Nucl. Phys. {\bf B249} (1985) 1.}
\bibitem{AKMRV}{D. Amati, K. Konishi, Y. Meurice, G. Rossi and G. Veneziano,
Phys. Rep. 162 (1988) 169.}
\bibitem{NSVZtwo}{V.A. Novikov, M.A. Shifman, A.I. Vainshtein and
V.I. Zakharov, Nucl. Phys. {\bf B260} (1985) 157.}

\bibitem{ADS}{ I. Affleck, M. Dine and N. Seiberg, Nucl. Phys. B241
(1984) 493; Nucl. Phys. {\bf B256} (1985) 557.  }


\bibitem{FS}{ J. Fuchs and M. G. Schmidt, Z. Phys. C30 (1986) 161. }

\bibitem{FP}{D. Finnell and P. Pouliot,
Nucl. Phys. B453 (1995) 225, {\tt hep-th/9503115}.}

\bibitem{KS}{A. Kovner and M. Shifman, Phys. Rev {\bf D 56} (1997) 2396,
{\tt hep-th/9702174}.}

\bibitem{SVrev}{ M. Shifman and A. Vainshtein, ``Instantons versus Supersymmetry:
Fifteen Years Later'', {\tt hep-th/9902018}.}

\bibitem{MO}{N. Dorey, V.V. Khoze and M.P. Mattis, 
Phys.~Rev.~D54 (1996) 2921, {\tt hep-th/9603136}; Phys.~Rev.~D54 (1996) 7832,
{\tt hep-th/9607202}.}

\bibitem{measone}{N. Dorey, V.V. Khoze and M.P. Mattis, 
Nucl. Phys. {\bf B513} (1998) 681,
{\tt hep-th/9708036}.}

\bibitem{meastwo}{N. Dorey, T.J. Hollowood,
V.V. Khoze and M.P. Mattis, Nucl. Phys. {\bf B519} (1998) 470,
{\tt hep-th/9709072}.}

\bibitem{KMS}{V.V. Khoze, M.P. Mattis and M.J. Slater, Nucl. Phys.
{\bf B536} (1998) 69, {\tt hep-th/9804009}.}

\bibitem{DHKMV}{N. Dorey, T.J. Hollowood, V.V. Khoze, M.P. Mattis and
S. Vandoren, ``Multi-Instantons and Maldacena's Conjecture'', 
{\tt hep-th/9810243};
``Multi-Instanton Calculus and the AdS/CFT Correspondence in $\N=4$
Superconformal Field Theory'', Nucl. Phys. {\bf B} ({\it to appear\/}),
{\tt hep-th/9901128}.}

\bibitem{SW}{ N. Seiberg and E. Witten, Nucl. Phys. {\bf B426} (1994) 19, {\tt hep-th/9407087}.}

\bibitem{Maldacena}{J. Maldacena, Adv. Theor. Math. Phys. {\bf 2} (1998) 231,
{\tt hep-th/9711200}.  }

\bibitem{BFST}{ A.A. Belavin, V.A. Fateeev, A.S. Schwarz and Y.S. Tyupkin, 
Phys. Lett. {\bf 83B} (1979) 317.}

\bibitem{FFS}{ V.A. Fateeev, I.V. Frolov and A.S. Schwarz, 
Nucl. Phys. {\bf B154} (1979) 1.}

\bibitem{BL}{ B. Berg and M. L{\"u}scher, 
Comm. Math. Phys. {69} (1979) 57.}

\bibitem{Osborn}{H. Osborn, Ann. Phys. {\bf 135} (1981) 373.}

\bibitem{CG}{E. Cohen and C. Gomez, Phys. Rev. Lett. {\bf 52} (1984) 237.}

\bibitem{Zh}{ A.R. Zhitnitsky
Nucl. Phys. {\bf B340} (1990) 56.}

\bibitem{DKMTV}{N. Dorey, V.V. Khoze, M.P. Mattis, D. Tong and
S. Vandoren,  Nucl. Phys. {\bf B502} (1997) 94,
{\tt hep-th/9703228}.}

\bibitem{GPY}{D.J. Gross, R.D. Pisarski and L.G. Yaffe, 
Rev. Mod. Phys. {\bf 53} (1981) 43.}

\bibitem{SWthree}{ N. Seiberg and E. Witten, ``Gauge Dynamics and 
Compactification to Three Dimensions'', {\tt hep-th/9607163}.}

\bibitem{Windx}{E. Witten, Nucl. Phys. {\bf B202} (1982) 253.}

\bibitem{BPST}{A. A. Belavin, A. M. Polyakov, A. Schwartz and Y. Tyupkin,
Phys. Lett. {\bf 59B} (1975) 85.}

\bibitem{HS}{B.J. Harrington and H.K. Shepard, Phys. Rev. {\bf D17} (1978) 2122.}


\bibitem{Nahmtwo}{W. Nahm, ``Self-dual Monopoles and Calorons'',
in: Lecture Notes in Physics 201, Eds. G. Denado {\it et. al.} 1984,
p. 189.}

\bibitem{Garland}{H. Garland and M.K. Murray, Comm. Math. Phys. {\bf 120} 
(1988) 335}.

\bibitem{thm}{G. 't Hooft, Nucl. Phys. {\bf B79} (1974) 276.}
\bibitem{polm}{A.M. Polyakov, JETP Lett. {\bf 20} (1974) 194.}

\bibitem{Bog}{ E.B. Bogomol'nyi, Sov. J. Phys {\bf 24} (1977) 97.}
\bibitem{PS}{ M.K. Prasad and C.M. Sommerfield, Phys. Rev. Lett.
{\bf 35} (1975) 760.}

\bibitem{Pol}{ A.M. Polyakov, Nucl. Phys. {\bf B120} (1977) 429. }

\bibitem{AHW}{ I. Affleck, J. Harvey and E. Witten, Nucl. Phys.
{\bf B206} (1982) 413.} 

\bibitem{PP}{ J. Polchinski and P. Pouliot, Phys. Rev. {\bf D56} (1997) 
6601, {\tt hep-th/9704029}.}

\bibitem{dkmthreed}{N. Dorey, V.V. Khoze and M.P. Mattis, 
Nucl. Phys. 
{\bf B502} (1997) 94,  {\tt hep-th/9704197}. }

\bibitem{Cal}{ C. Callias, Comm. Math. Phys. {\bf 62} (1978) 213.}

\bibitem{Brodie:1998bv}
J.~Brodie, Nucl. Phys. {\bf B532} (1998) 137,
{\tt hep-th/9803140}.

\bibitem{Witpbr}{E. Witten, 
Nucl. Phys. {\bf B460} (1996) 335, 
{\tt hep-th/9510135}.}

\bibitem{Polch}{J. Polchinski, ``TASI Lectures on D-Branes'', {\tt
hep-th/9611050}.}

\bibitem{EGK}{S. Elitzur, A. Giveon and D. Kutasov, 
 Phys. Lett. {\bf B400}
(1997) 269, {\tt hep-th/9702014}.}

\bibitem{Witqcd}{E. Witten,
Nucl. Phys. {\bf B507} (1997) 658, 
{\tt hep-th/9706109}.}

\bibitem{Adda}{ A. D'Adda and P. Di Vecchia, 
Phys. Lett. {\bf 73B} (1978) 162.}

\bibitem{Seiberg93}
{N. Seiberg, Phys. Lett. {\bf 318B} (1993) 469, {\tt hep-th/9309335}.}

\end{thebibliography}
\end{document}